\documentclass[prd,showpacs,altaffilletter,twocolumn,
superscriptaddress,floatfix,nofootinbib,preprintnumbers]{revtex4-1} 

\pdfoutput=1
\usepackage{amsfonts,amsmath,graphicx,bm} 
\usepackage{float}
\usepackage{dcolumn}
\usepackage[pdftex,
       colorlinks=true,
       urlcolor=blue,       
       linkcolor=red,       
       pdfpagemode=None,
       bookmarksopen=true]{hyperref}
\usepackage[capitalise]{cleveref}
\usepackage{ifpdf}
\usepackage{multirow}
\usepackage[colorlinks=true]{hyperref}
\usepackage{url}
\usepackage{slashed}

\newcommand{\glasgow}{SUPA, School of Physics and Astronomy,
  University of Glasgow, Glasgow, G12 8QQ, UK}

\def\today{\number\day\space\ifcase\month\or
January\or February\or March\or April\or May\or June\or
July\or August\or September\or October\or November\or December\fi
\space\number\year}
\newcount\mins \newcount\hours
\def\now{\hours=\time \mins=\time
	\divide\hours by60 \multiply\hours by60 \advance\mins by-\hours
	\divide\hours by60 
	\number\hours:\ifnum\mins<10 0\fi\number\mins }

\newcommand{\schionep}{0.720(34)\times 10^{-2}}
\newcommand{\schionem}{1.161(54)\times 10^{-2}}
\newcommand{\schizm}{2.374(33)\times 10^{-2}}
\newcommand{\schizp}{0.609(14)\times 10^{-2}}
\newcommand{\schiT}{0.891(44)\times 10^{-2}}
\newcommand{\schiAT}{0.441(33)\times 10^{-2}}

\newcommand{\schionepMh}{4.06(20)\times 10^{-4}\mathrm{GeV}^{-2}}
\newcommand{\schionemMh}{6.55(31)\times 10^{-4}\mathrm{GeV}^{-2}}
\newcommand{\schiTMh}{5.03(25)\times 10^{-4}\mathrm{GeV}^{-2}}
\newcommand{\schiATMh}{2.49(19)\times 10^{-4}\mathrm{GeV}^{-2}}

\begin{document}

\title{$\bar{b}c$ susceptibilities from fully relativistic lattice QCD}

\author{Judd \surname{Harrison}}
\email[]{judd.harrison@glasgow.ac.uk}
\affiliation{\glasgow}

\collaboration{HPQCD Collaboration}
\email[]{http://www.physics.gla.ac.uk/HPQCD}


\begin{abstract}
We compute the $\bar{h}c$ (pseudo)scalar, (axial-)vector and (axial-)tensor susceptibilities as a function of $u=m_c/m_h$ between $u=m_c/m_b$ and $u=0.8$ using fully relativistic lattice QCD, employing nonperturbative current renormalisation and using the second generation 2+1+1 MILC HISQ gluon field configurations. We include ensembles with $a\approx 0.09\mathrm{fm}$, $0.06\mathrm{fm}$, $0.045\mathrm{fm}$ and $0.033\mathrm{fm}$ and we are able to reach the physical $b$-quark on the two finest ensembles. At the physical $m_h=m_b$ point we find $\overline{m}_b^2 \chi_{1^+}=\schionep$, $\overline{m}_b^2 \chi_{1^-}=\schionem$, $\chi_{0^-}=\schizm$, $\chi_{0^+}=\schizp$. Our results for the (pseudo)scalar, vector and axial-vector are compatible with the expected small size of nonperturbative effects at $u=m_c/m_b$. We also give the first nonperturbative determination of the tensor susceptibilities, finding $\overline{m}_b^2 \chi_{T}=\schiT$ and $\overline{m}_b^2 \chi_{AT}=\schiAT$. Our value of $\overline{m}_b^2\chi_{AT}$ is in good agreement with the $\mathcal{O}(\alpha_s)$ perturbation theory, while our result for $\overline{m}_b^2\chi_{T}$ is in tension with the $\mathcal{O}(\alpha_s)$ perturbation theory at the level of $2\sigma$. These results will allow for dispersively bounded parameterisations to be employed using lattice inputs for the full set of $h\to c$ semileptonic form factors in future calculations, for heavy-quark masses in the range $1.25\times m_c \leq m_h \leq m_b$.
\end{abstract}

\maketitle



\section{Introduction}
\label{sec:intro}
Lattice QCD studies of the semileptonic decays of $B_{(s,c)}$-mesons to vector-mesons via the $b\to c\ell\bar{\nu}$ weak transition have progressed significantly in recent years, with lattice form factor results becoming available away from zero recoil for $B\to D^*\ell\bar{\nu}$~\cite{FermilabLattice:2021cdg,Harrison:2023dzh,Aoki:2023qpa},  $B_s\to D_s^{(*)}\ell\bar{\nu}$~\cite{Harrison:2021tol,EuanBsDs} and  $B_c\to J/\psi\ell\bar{\nu}$~\cite{Harrison:2020gvo}. However, lattice predictions for the differential decay rate for $B\to D^*\ell\bar{\nu}$ have been found to be in tension with that measured by the Belle experiment~\cite{Belle:2018ezy}. Moreover, predictions for the ratios of form factors obtained by combining earlier zero-recoil lattice results with light-cone sum-rules~(LCSR) and QCD sum-rules~(QCDSR) using the heavy-quark expansion~(HQE) through order $\mathcal{O}(1/m_b,1/m_c^2)$~\cite{Bordone:2019guc} show some disagreement with the more recent lattice-only results.

For fully relativistic lattice calculations, it is typical to compute form factors at multiple heavy-quark masses, $m_h$, below and ranging up to $m_b$, in order to control discretisation effects appearing as powers of $(am_h)^2$~\cite{Aoki:2023qpa,Harrison:2023dzh,Harrison:2021tol,Harrison:2020gvo,EuanBsDs}. The lattice data is then fit using a function chosen to describe both the physical heavy mass dependence and kinematics, as well as discretisation and quark mass mistuning effects. The choice of this fit function is one potential origin of the discrepancy seen between lattice-only results and the results combining LCSR, zero-recoil lattice and HQE for $B\to D^*$. 

In the continuum, the $B\to D^{(*)}$ form factors obey dispersive bounds and may be described using the Boyd-Grinstein-Lebed~(BGL) parameterisation~\cite{Boyd:1997kz}, which we briefly describe below. This parameterisation is formulated using the variable 
\begin{align} 
z(q^2,t_+,t_0)=\frac{\sqrt{t_+-q^2}-\sqrt{t_+-t_0}}{\sqrt{t_+-q^2}+\sqrt{t_+-t_0}},
\end{align} 
where $t_+$ is the $\bar{b}c$ two particle production threshold for the relevant current and $t_0$ is a free parameter which may be chosen between $t_+$ and $-\infty$. $z$ maps the physical $q^2$ region to within the unit circle and the branch cut $q^2\geq t_+$ to the unit circle. The susceptibilities, $\chi_{J^P}$, are defined in terms of the two point correlation functions of $\bar{b}c$ currents with quantum numbers $J^P$ (see~\cref{sec:theory}), and are typically computed using perturbation theory. The susceptibilities can then be related via the optical theorem and crossing symmetry to a sum over the squared magnitudes of exclusive hadronic matrix elements. Because each contribution in this sum is positive semidefinite, the sum may be restricted to just the lowest two particle contribution, corresponding to $B\to D^{(*)}$. This results in inequalities involving the helicity-basis form factors, $F$, integrated over the unit circle in $z$. These inequalities take the form
\begin{align}\label{ineqbgl}
\int_{\mathcal{C}} |P(z)\phi(z) F(z)|^2 \leq 1,
\end{align}
where $\phi(z)$, referred to as outer functions, are analytic functions on the open unit disk, which also absorb a factor of $1/\sqrt{\chi_{J^P}}$ in order to set the right-hand side of~\cref{ineqbgl} to unity. The Blashke factors, $P(z)$, have magnitude 1 on the unit circle, and remove subthreshold poles appearing in the form factor. $P(z)\phi(z) F(z)$ can then be analytically continued to real $z$ corresponding to the physical semileptonic region of $q^2$. Because $P(z)\phi(z) F(z)$ is analytic on the open unit disc, it may be expanded as a polynomial in $z$ as
\begin{align}
P(z)\phi(z) F(z)=\sum_{n=0}^\infty a_n z^n,
\end{align}
resulting in the standard BGL parameterisation for the form factor,
\begin{align}\label{BGLparam}
F(z)=\frac{1}{P(z)\phi(z)}\sum_{n=0}^\infty a_n z^n,
\end{align}
where from~\cref{ineqbgl} the coefficients satisfy the inequality
\begin{align}
\sum_{n=0}^\infty |a_n|^2 \leq 1.
\end{align}
Note that stronger dispersive bounds than those of the original BGL approach may be formulated by decomposing the polarisation tensor for a given current in terms of a full set of virtual vector boson polarisation vectors~\cite{Gubernari:2023puw}. Also note that the BGL parameterisation corresponds to the special case where the lowest two particle threshold, $t_+$, corresponds to the production threshold, $t_\Gamma$, of the initial and final state mesons for the form factors of interest. In the more general case where $t_\Gamma \geq t_+$, the integral in~\cref{ineqbgl} is restricted to an arc on the unit circle, and instead of a simple sum of powers of $z$ as in~\cref{BGLparam}, one finds a sum over polynomials in $z$ constructed to be orthonormal on the corresponding arc~\cite{Gubernari:2023puw}.

On the lattice, the HPQCD collaboration has previously employed two different fit functions to reach the physical continuum. Earlier works on $B_s\to D_s^*$ and $B_c\to J/\psi$ used a `pseudo-BGL' fit~\cite{Harrison:2021tol,Harrison:2020gvo}, where a power series in the conformal variable $z$ was used to describe the kinematic dependence of the form factors in the QCD basis, together with a term describing the $\bar{b}c$ subthreshold poles. However, these fits omitted the outer functions of the full BGL parameterisation~\cref{BGLparam}. More recently, for a combined analysis of $B\to D^*$ and $B_s\to D_s^*$, the HPQCD collaboration used a fit to the HQET form factors using a simple power series in $w-1$, choosing priors to ensure the continuum BGL coefficients were not significantly constrained relative to the unitarity bounds~\cite{Harrison:2023dzh}. In both cases, coefficients included $(\Lambda/m_h)^i$ corrections encoding the physical heavy mass dependence. 

Neither of these fit functions is ideal. The pseudo-BGL fit neglects the dependence on the heavy-quark mass of the outer functions, as well as losing the ability to choose prior widths informed by the unitarity constraints. On the other hand, the HQET fit includes limited information about the known pole structure of the form factors with varying heavy quark mass. Ideally a full BGL fit would be used to fit lattice data, augmenting the BGL coefficients with $\Lambda/m_h$ terms to describe the dependence of the lattice data on heavy-quark mass while using lattice inputs to describe the subthreshold pole masses and susceptibilities. This approach is complicated by the susceptibilities, which determine the overall normalisation of the outer functions~\cite{Boyd:1997kz}. The susceptibilities for the $\bar{b}c$ (pseudo)scalar and (axial-)vector currents are known perturbatively for the physical $b$-quark to 3-loops~\cite{Hoff:2011ge,Grigo:2012ji}, with nonperturbative condensate contributions expected to be extremely small. These susceptibilities have also recently been computed nonperturbatively using lattice QCD~\cite{Martinelli:2021frl}, where surprising tension at the level of $\approx 2\sigma$ was found between the lattice and perturbation theory at the $m_h=m_b$ point. The tension is particularly surprising because of the good consistency seen between the continuum perturbation theory and the equivalent heavyonium quantities~\cite{HPQCD:2008kxl,PhysRevD.91.054508}.

Recently, lattice form factor calculations have also been extended to include the tensor form factors needed to analyse and constrain new physics~\cite{Parrott:2022rgu,Harrison:2023dzh}. Dispersive parameterisations of the tensor form factors require tensor susceptibilities computed from the polarisation tensor of the corresponding tensor currents. For $\bar{b}c$ currents, these are currently only available from perturbation theory to $\mathcal{O}(\alpha_s)$~\cite{Bharucha:2010im}.

In this work, we compute the full set of (pseudo)scalar, (axial-)vector and (axial-)tensor susceptibilities as a function of $u=m_c/m_h$ between $u^\mathrm{phys}=m_c/m_b$ and $\approx 0.8$ using the $a\approx 0.09\mathrm{fm}$, $0.06\mathrm{fm}$, $0.045\mathrm{fm}$ and $0.03\mathrm{fm}$ second generation MILC HISQ 2+1+1 gauge field ensembles. This will provide an additional check of the perturbation theory and lattice results~\cite{Martinelli:2021frl} for the (pseudo)scalar and (axial-)vector susceptibilities, as well as providing new lattice results for the $\bar{b}c$ (axial-)tensor susceptibilities. These new (axial-)tensor susceptibilities will allow future heavy-HISQ calculations of form factors for exclusive $b\to c$ processes to use the full dispersive parameterisation for all form factors, while using lattice results for all inputs. This calculation will also lead to a future calculation of the heavy-light susceptibilities, where nonperturbative condensate contributions are expected to be more sizeable.

\section{Theoretical Background}
\label{sec:theory}
\subsection{(Pseudo)scalar and (Axial-)vector Currents}
The susceptibilities are related to polarisation functions, which are decomposed according to Lorentz structure, and are defined in terms of current-current correlators by
\begin{align}\label{polfuncvec}
(-q^2g_{\mu\nu}+q_\mu q_\nu)\Pi^\delta(q^2)+q_\mu q_\nu\Pi^\delta_L(q^2)&\nonumber\\
=i\int dx e^{iq x} \langle 0| T &j_\mu^\delta(x)j_\nu^{\delta\dagger}(0)|0 \rangle,
\end{align}
for the vector and axial-vector currents $j^V_\mu=\bar{\psi}_h\gamma_\mu\psi_c$, $j^A_\mu=\bar{\psi}_h\gamma_\mu\gamma^5\psi_c$ and by
\begin{align}\label{polfuncscal}
q^2\Pi^\delta(q^2)=i\int dx^4 e^{iq x} \langle 0| T &j^\delta(x)j^{\delta\dagger}(0)|0 \rangle,
\end{align}
for the scalar and pseudoscalar currents $j^S=\bar{\psi}_h\psi_c$ and $j^P=i\bar{\psi}_h\gamma^5\psi_c$.
Moments of the heavy-light current correlators were computed up to three loops in perturbation theory in~\cite{Hoff:2011ge,Grigo:2012ji} in the $\overline{\mathrm{MS}}$ scheme. The three-loop $\overline{\mathrm{MS}}$ results for the limit $q^2\to 0$ are expressed as 
\begin{align}\label{pertdef}
\bar{\Pi}^\delta(q^2)=\frac{3}{16\pi^2}\sum_{n\ge -1} \bar{C}_n^\delta(u) z^n
\end{align}
where $u=m_c/m_h$ and $z=q^2/m_h^2$. The susceptibilities are then defined at the point $q^2=0$ by
\begin{align}\label{suscdef}
\chi_{1^+}(q^2=0)&\equiv\frac{1}{2}\frac{\partial^2}{\partial^2 q^2} \left(q^2 {\Pi}^A(q^2)\right)\Big|_{q^2=0},\nonumber\\
\chi_{1^-}(q^2=0)&\equiv\frac{1}{2}\frac{\partial^2}{\partial^2 q^2} \left(q^2 {\Pi}^V(q^2)\right)\Big|_{q^2=0},\nonumber\\
\chi_{0^-}(q^2=0)&\equiv\frac{(m_h+m_c)^2}{2}\frac{\partial^2}{\partial^2 q^2} \left(q^2 {\Pi}^P(q^2)\right)\Big|_{q^2=0},\nonumber\\
\chi_{0^+}(q^2=0)&\equiv\frac{(m_h-m_c)^2}{2}\frac{\partial^2}{\partial^2 q^2} \left(q^2 {\Pi}^S(q^2)\right)\Big|_{q^2=0},
\end{align}
where in the final two lines we have used the partially conserved axial-vector and vector current relations. Inserting \cref{pertdef} into \cref{suscdef} gives the susceptibilities in terms of the perturbatively computed moments of~\cite{Hoff:2011ge,Grigo:2012ji} as
\begin{align}\label{pertchi}
\chi_{1^+}(q^2=0)&=\frac{3}{\overline{m}_h^2 16\pi^2}\bar{C}_1^A(u) \nonumber\\
\chi_{1^-}(q^2=0)&=\frac{3}{\overline{m}_h^2 16\pi^2}\bar{C}_1^V(u) \nonumber\\
\chi_{0^-}(q^2=0)&={(1+u)^2}\frac{3}{16\pi^2}\bar{C}_1^P(u)\nonumber\\
\chi_{0^+}(q^2=0)&={(1-u)^2}\frac{3}{16\pi^2}\bar{C}_1^S(u).
\end{align}
where the $\bar{C}_1^\delta$ are given by~\cite{Hoff:2011ge,Grigo:2012ji} 
\begin{align}\label{Coneu}
\bar{C}_1^\delta(u)=\bar{C}_1^{(0),\delta}(u)+\frac{\alpha_s}{\pi}\bar{C}_1^{(1),\delta}(u)+\left(\frac{\alpha_s}{\pi}\right)^2\bar{C}_1^{(2),\delta}(u)
\end{align}
with $\alpha_s=\alpha_s(\mu)$ and $\mu=\overline{m}_h(\overline{m}_h)$. Note that~\cref{Coneu} does not include nonperturbative condensate contributions. To set $\overline{m}_h$ we use $\overline{m}_c(3\mathrm{GeV})=0.9858(51)\mathrm{GeV}$ from~\cite{Hatton:2020qhk} to compute $\overline{m}_h(3\mathrm{GeV})=\overline{m}_c(3\mathrm{GeV})/u$, which we then run to $\overline{m}_h(\overline{m}_h)$. We use $\alpha_{\overline{\mathrm{MS}}}(5\mathrm{GeV},n_f=4)=0.2128(25)$ from~\cite{PhysRevD.91.054508}, together with the 4-loop running~\cite{vanRitbergen:1997va}. We use $u^\mathrm{phys}=1/4.578(12)$ computed in pure QCD from~\cite{Hatton:2021syc}. We include an uncertainty for the three-loop result of $\sigma_3(\alpha_s/\pi)^3$ where $\sigma_3$ is the root-mean-square of $\bar{C}_1^{(0),\delta}(u)$, $\bar{C}_1^{(1),\delta}(u)$ and $\bar{C}_1^{(2),\delta}(u)$. 

\subsection{(Axial-)Tensor Currents}
The susceptibilities are defined analogously for the tensor and axial-tensor currents, with one of the tensor indices contracted with $q^\alpha$
\begin{align}\label{tensorcurrents}
j^T_\mu=\bar{\psi}_h\sigma_{\mu\alpha}q^\alpha\psi_c,~~~~~~~j^{AT}_\mu=\bar{\psi}_h\sigma_{\mu\alpha}\gamma^5 q^\alpha\psi_c.
\end{align}
The polarisation functions for the (axial-)tensor currents given in~\cref{tensorcurrents} are defined by
\begin{align}\label{polfuncvec}
\left(q_\mu q_\alpha-q^2g_{\mu\nu}\right)\Pi^{T}(q^2)&=\nonumber\\
i\int dx e^{iq x}& \langle 0| T j_\mu^T(x)j_\nu^{T\dagger}(0)|0 \rangle,\nonumber\\
\left(q_\mu q_\nu-q^2g_{\mu\nu}\right)\Pi^{AT}(q^2)&=\nonumber\\
i\int dx e^{iq x}& \langle 0| T j_\mu^{AT}(x)j_\nu^{AT\dagger}(0)|0 \rangle.\nonumber\\
\end{align}
Note that because $\sigma_{\mu\nu}$ is antisymmetric, there is no longitudinal piece proportional to the projector $q_\mu q_\nu/q^2$.
The (axial-)tensor polarisation functions require 3 subtractions~\cite{Bharucha:2010im}, and the susceptibilities are defined by
\begin{align}\label{suscdefT}
\chi^{J=1}_{(A)T}(q^2=0)&\equiv\frac{1}{6}\frac{\partial^3}{\partial^3 q^2} \left(q^2 {\Pi}^{(A)T}(q^2)\right)\Big|_{q^2=0}.
\end{align}
Since the $J=0$ components of the polarisation tensors are identically zero, we will omit the $J=1$ label from the susceptibilities and write $\chi_{(A)T}\equiv \chi^{J=1}_{(A)T}$ from now on. The tensor currents also require renormalisation in the continuum. This is typically performed in the $\overline{\mathrm{MS}}$ scheme, and the tensor susceptibilities are dependent upon the renormalisation scale $\mu$.

\section{Lattice Calculation}\label{sec:lattcalc}
Following~\cite{Martinelli:2021frl}, the continuum Euclidean correlation functions that we wish to compute are 
\begin{align}
C_{A}(t)=&\frac{Z_V^2}{3}\sum_{j=1}^3\int dx^3 \langle 0|T \bar{h}\gamma_j^E\gamma_5^E c(x) \bar{c}\gamma_j^E\gamma_5^E h(0)|0\rangle,\nonumber\\
C_{V}(t)=&\frac{Z_V^2}{3}\sum_{j=1}^3\int dx^3 \langle 0|T \bar{h}\gamma_j^E c(x) \bar{c}\gamma_j^E h(0)|0\rangle,\nonumber\\
C_{P}(t)=&\int dx^3 \langle 0|T \bar{h}\gamma_5^E c(x) \bar{c}\gamma_5^E h(0)|0\rangle,\nonumber\\
C_{S}(t)=&\int dx^3 \langle 0|T \bar{h}c(x) \bar{c}h(0)|0\rangle,\nonumber\\
C_{T}(t)=&\frac{Z_T^2}{3}\sum_{j=1}^3\int dx^3 \langle 0|T \bar{h}\sigma_{j0}^E c(x) \bar{c}\sigma_{j0}^E h(0)|0\rangle,\nonumber\\
C_{AT}(t)=&\frac{Z_T^2}{3}\sum_{j=1}^3\int dx^3 \langle 0|T \bar{h}\sigma_{j0}^E\gamma_5^E c(x) \bar{c}\sigma_{j0}^E\gamma_5^E h(0)|0\rangle,
\end{align}
where $\gamma_j^E$ and $\gamma_5^E$ are Euclidean gamma matrices and for the (axial-)vector and (axial-)tensor currents we require the additional current renormalisation factors $Z_V$ and $Z_T$ respectively.

Using the definitions of the susceptibilities, together with the definitions of the polarisation functions, the susceptibilities may be expressed in terms of these correlation functions as~\cite{Martinelli:2021frl}
\begin{align}\label{moments}
\chi_{1^+}(q^2=0)&=\frac{1}{12}\int dt ~t^4 C_{A}(t),\nonumber\\
\chi_{1^-}(q^2=0)&=\frac{1}{12}\int dt ~t^4 C_{V}(t),\nonumber\\
\chi_{0^-}(q^2=0)&=\frac{1}{12}(m_h+m_c)^2\int dt ~t^4 C_{P}(t),\nonumber\\
\chi_{0^+}(q^2=0)&=\frac{1}{12}(m_h-m_c)^2\int dt ~t^4 C_{S}(t),\nonumber\\
\chi_{(A)T}(q^2=0)&=\frac{1}{12}\int dt ~t^4 C_{(A)T}(t).
\end{align}
We compute the required correlation functions using the HISQ~\cite{PhysRevD.75.054502} formalism for the $h$ and $c$ quarks on the MILC 2+1+1 HISQ gluon field configurations detailed in~\cref{tab:gaugeinfo}. We use the local spin-taste operators $1\otimes 1$, $\gamma_5\otimes \gamma_5$, $\gamma_j\otimes \gamma_j$ and $\gamma_j\gamma_5\otimes \gamma_j\gamma_5$ for the $S$, $P$, $V$ and $A$ currents respectively. For the tensor currents $T$ and $AT$ we use $\gamma_j\gamma_0\otimes \gamma_j\gamma_0$ and $\gamma_i\gamma_k\otimes \gamma_i\gamma_k$ respectively, with $i$ and $k$ chosen as spatial directions and $i\neq k$. Note that we use the local currents to avoid tree-level discretisation errors. The valence charm and heavy-quark masses used in this work are given in~\cref{tab:valmasses}, with $am_h=0.9\approx am_b$ on set 3 and $am_h=0.625\approx am_b$ on set 4. Note that because we use the HISQ formalism for both heavy and charm quarks, we can use $u=m_c^\mathrm{val}/m_h^\mathrm{val}$ directly. The ensembles we use include physically tuned charm and strange quarks in the sea, as well as unphysically heavy light sea quarks on sets $1-4$. While the effect of using heavier-than-physical light quarks is expected to be very small, we also include a single ensemble, set 5, with physically tuned light quarks, in order to constrain these effects.

\begin{table}
\caption{Details of the gauge field configurations used in our calculation \cite{PhysRevD.87.054505,PhysRevD.82.074501}. We use the Wilson flow parameter~\cite{Borsanyi:2012zs}, $w_0$, to fix 
the lattice spacing given in column 2. The physical value of $w_0$ was determined in \cite{PhysRevD.88.074504} to be 0.1715(9)fm and the values of $w_0/a$, 
which are used together with $w_0$ to compute $a$, were taken from \cite{PhysRevD.96.034516,PhysRevD.91.054508,EuanBsDs}. Set 1 with $w_0/a=1.9006(20)$ is referred to as `fine', set 2 with $w_0/a=2.896(6)$ as `superfine', set 3 with $w_0/a=3.892(12)$ as `ultrafine' and set 4 with $w_0/a=1.9518(7)$ as `physical fine'. $n_\mathrm{cfg}$ is the number of configurations that we use here. $am_{l0}$, $am_{s0}$ and $am_{c0}$ are the masses of the sea up/down, strange and charm quarks in lattice units. We also include the approximate mass of the Goldstone pion, computed in~\cite{Bazavov:2017lyh}.\label{tab:gaugeinfo}}
\begin{tabular}{c c c c c c c c c}\hline
 Set &$a$ & $L_x\times L_t$ &$am_{l0}$&$am_{s0}$& $am_{c0}$ & $M_\pi$ &$n_\mathrm{cfg}$ \\ 
  & $(\mathrm{fm})$&  &&&  & $(\mathrm{MeV})$ & \\ \hline
1 & $0.0902$   & $32\times 96 $    &$0.0074$ &$0.037$  & $0.440$ & $316$ & $1000$\\
2 & $0.0592$   & $48\times 144  $    &$0.0048$ &$0.024$  & $0.286$ & $329$ & $500$\\
3 & $0.0441$   &$ 64\times 192  $    &$0.00316$ &$0.0158$  & $0.188$ & $315$ & $375$\\
4 & $0.0327$  &$ 96\times 288  $    &$0.00223$ &$0.01115$  & $0.1316$ & $309$ & $100$\\
5 & $0.0879$  &$ 64\times 96  $    &$0.0012$ &$0.0363$  & $0.432$ & $129$ & $500$\\\hline
\end{tabular}
\end{table}

\begin{table}
\centering
\caption{Details of the charm and heavy valence masses. \label{tab:valmasses}}
\begin{tabular}{c c c}\hline
 Set & $am_h^\mathrm{val}$  & $am_c^\mathrm{val} $\\ \hline
1 & $0.5,0.55,0.6,0.65,0.7,0.75,0.8$    & $0.449$  \\\hline
2 & $0.427,0.525,0.55,0.6,0.65,0.7,0.75,0.8$    & $0.274$   \\\hline
3 &  \begin{tabular}{@{}c@{}}$0.25,0.3,0.35,0.4,0.45,0.5,0.55,0.6,0.65,$  \\ $0.7,0.75,0.8,0.85,0.9$\end{tabular}   & $0.194$  \\\hline
4 & $0.2$,$0.25$,$0.3$,$0.45$,$0.625$   & $0.137$ \\ \hline
5 & $0.5,0.55,0.6,0.65,0.7,0.75,0.8$   & $0.433$ \\ \hline
\end{tabular}
\end{table}

\begin{table}
\centering
\caption{The second column gives the values of $Z_V(\mu=2\mathrm{GeV})$ at zero valence quark mass computed in~\cite{Hatton:2019gha} and~\cite{Hatton:2020qhk} in the RI-SMOM scheme. Note that $Z_V$ on set 5 is equal to that on set 1. The third column gives the values of $Z_T(\mu=4.8~\mathrm{GeV})$ from~\cite{Hatton:2020vzp} for the tensor operators used in this work. Note that since~\cite{Hatton:2020vzp} did not include set 4, we use a value here obtained by extrapolating the other values in $a^2$ as described in the text. \label{tab:Z}}
\begin{tabular}{c c c }\hline
 Set & $Z_V(\mu=2\mathrm{GeV})$  & $Z_T(\mu=4.8~\mathrm{GeV})$ \\ \hline
1 & $0.98445(11)$&$1.0029(43)$  \\\hline
2 &  $0.99090(36)$&$1.0342(43)$   \\\hline
3 &  $0.99203(108)$&$1.0476(42)$  \\\hline
4 &  $0.99296(21)$&$1.0570(50)$ \\ \hline
5 &  $0.98445(11)$&$1.0029(43)$ \\ \hline
\end{tabular}
\end{table}

The $Z_V$ factors for the local vector current were computed in~\cite{Hatton:2019gha} and~\cite{Hatton:2020qhk}, extrapolated to zero valence quark mass. For the tensor, we use the results of~\cite{Hatton:2020vzp}, which used an intermediate RI-SMOM scheme to match the lattice tensor current to the continuum tensor current in the $\overline{\mathrm{MS}}$ scheme. We use the values computed using a matching scale of $\mu=2\mathrm{GeV}$ which we subsequently run to $\overline{m}_h(\overline{m}_h)$ using the 3-loop anomalous dimension~\cite{Gracey:2000am}. For HISQ, chiral symmetry means that the local vector and tensor currents used here have the same renormalisation factors in the zero valence quark mass limit as their axial counterparts to all orders in perturbation theory~\cite{Sharpe:1993ur}. It was shown in~\cite{Hatton:2019gha} that $Z_V$ computed using the RI-SMOM scheme is free from condensate contamination, while $Z_T$ includes a correction to remove condensate contributions explicitly~\cite{Hatton:2020vzp}. We may therefore use $Z_A=Z_V$ and $Z_{AT}=Z_T$, which will differ only by discretisation effects, and so give the correct continuum limit. The values of $Z_V$ and $Z_T$ used here are given in~\cref{tab:Z}. For each value of $am_h$ on each ensemble, we run $Z_T$ to the $\overline{\mathrm{MS}}$ mass $\overline{m}_h(\mu=\overline{m}_h)$ which we determine using the physical value of $\overline{m}_c(3\mathrm{GeV})=0.9858(51)\mathrm{GeV}$ from~\cite{Hatton:2020qhk} together with the ratio of lattice masses
\begin{equation}\label{mhbarmhbar}
\overline{m}_h(3\mathrm{GeV})=\overline{m}_c(3\mathrm{GeV})/u=\overline{m}_c(3\mathrm{GeV})/(am_c/am_h).
\end{equation}
Note that since~\cite{Hatton:2020vzp} did not include set 4, we use a value here obtained by extrapolating the other values. Following~\cite{Hatton:2020qhk}, we fit the condensate-corrected tensor renormalisation factors, at scale $\mu=2\mathrm{GeV}$, using the simple fit function
\begin{align}
Z_T(a,\mu=2\mathrm{GeV}) =  \sum_{i=0}^{i=4}\left(c_i + \sum_{j=1}^{j=3} b_{ij}\left(\frac{a\mu}{\pi}\right)^{2j}\right)\alpha_s(\pi/a)^i
\end{align}
taking priors of $0\pm 2$ for the coefficients $c_i$ and $b_{ij}$. Varying either $\mu$ or the lattice scale, $\pi/a$, by $\pm 50\%$ has a negligible effect on the extrapolated value, as does increasing the maximum order that we sum to in $i$ or $j$. Note that we neglect the statistical correlations between $Z_V$ and $Z_T$ as well as between the current renormalisation factors and the lattice data generated in this work.

\begin{figure}
\centering
\includegraphics[scale=0.3]{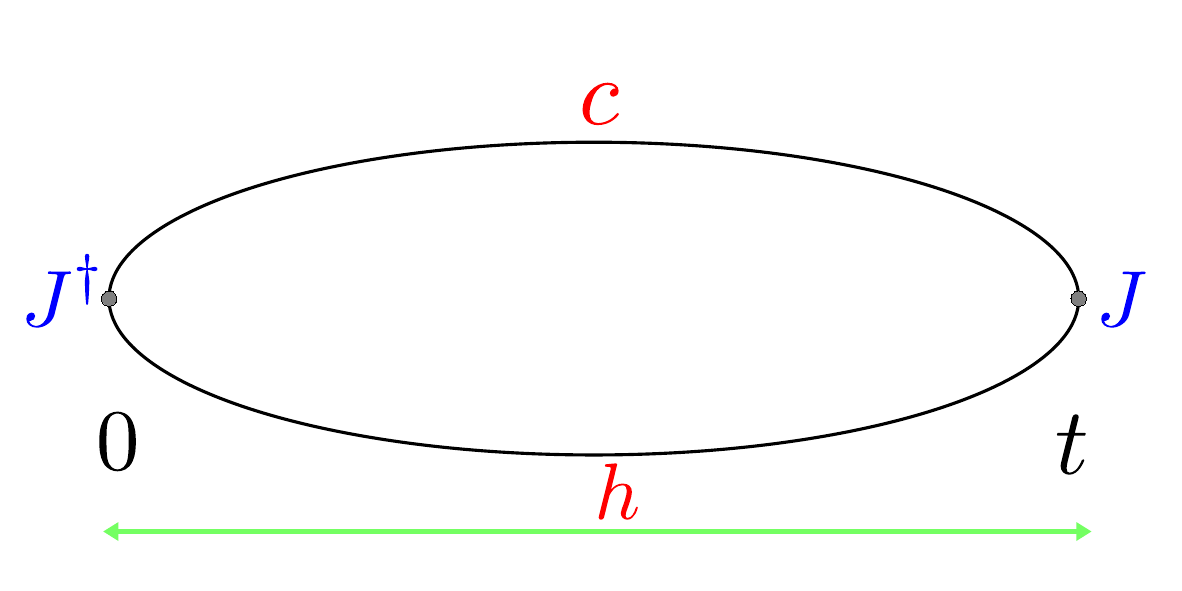}
\caption{\label{c2ptfig}Arrangement of heavy and charm quark propagators for currents $J=j^S,~j^P,~j^V_\mu,~j^A_\mu,~j^{T}_\mu,~j^{AT}_\mu$ defined in~\cref{sec:theory}.}
\end{figure}

We use random wall sources to increase statistical precision. The arrangement of propagators appearing in the correlation functions which we compute are shown in~\cref{c2ptfig}.
In terms of the staggered fields they are given by
\begin{align}
C^\mathrm{latt}_\delta=\frac{1}{L_x^3}\sum_{x;y,y'} \beta_\delta(x)& g^c_{ab}(t,x;0,y)\beta_\delta(y)\xi_{bc}(y)\nonumber\\
&\times \left( g^h_{ad}(t,x;0,y') \xi_{dc}(y')\right)^\ast
\end{align}
where $g^q$ is the staggered propagator for flavour $q$ and the random wall $\xi$ satisfies $\overline{\xi_{ac}(y)\xi^\ast_{bc}(y')}=\delta_{y,y'}\delta_{ab}$. $\beta_\delta(x)$ is the $x$-dependent phase factor corresponding to the local spin-taste operator in the staggered formalism.

The correlation functions we compute are periodic in time, and so we average $C^\mathrm{latt}(t)$ and $C^\mathrm{latt}(L_t-t)$ for $0<t<L_t/2$. We compute the time moments in~\cref{moments} on the lattice as
\begin{align}\label{lattmoments}
\chi_{1^+}^\mathrm{latt}(q^2=0)&=\frac{1}{12}\sum_{t=0}^{L_T/2} ~t^4 C^\mathrm{latt}_{A}(t),\nonumber\\
\chi_{1^-}^\mathrm{latt}(q^2=0)&=\frac{1}{12}\sum_{t=0}^{L_T/2} ~t^4 C^\mathrm{latt}_{V}(t),\nonumber\\
\chi_{0^-}^\mathrm{latt}(q^2=0)&=\frac{1}{12}(m_h+m_c)^2\sum_{t=0}^{L_T/2} ~t^4 C^\mathrm{latt}_{P}(t),\nonumber\\
\chi_{0^+}^\mathrm{latt}(q^2=0)&=\frac{1}{12}(m_h-m_c)^2\sum_{t=0}^{L_T/2} ~t^4 C^\mathrm{latt}_{S}(t),\nonumber\\
\chi_{(A)T}^\mathrm{latt}(q^2=0)&=\frac{1}{12}\sum_{t=0}^{L_T/2} ~t^4 C^\mathrm{latt}_{(A)T}(t).
\end{align}
The resulting values of $\chi_{\delta}^\mathrm{latt}(q^2=0)$ for the (pseudo)scalar, (axial-)vector and (axial-)tensor susceptibilities are given in~\cref{sec:applattdat}. The susceptibilities on a given ensemble are computed including all statistical correlations, which are then included in our subsequent chiral continuum extrapolation.

\section{Continuum Extrapolation}\label{sec:contextrap}
In order to reach the continuum we fit the lattice susceptibilities against a form including dependence on $u$, $am_h$, $am_c$ and the quark mass mistunings. We use the fit functions
\begin{align}\label{continuumfitfunc}
\chi^\mathrm{latt}_{1^\pm}&=\frac{1}{\overline{m}_h(\overline{m}_h)^2}\sum_{n=0}^{12} \bar{a}_{n}^{1^\pm}(1-u)^n \Delta_n^{1^\pm}\mathcal{N}^{1^\pm}_n, \nonumber\\
\chi^\mathrm{latt}_{(A)T}&=\frac{1}{\overline{m}_h(\overline{m}_h)^2}\sum_{n=0}^{12} \bar{a}_{n}^{(A)T}(1-u)^n \Delta_n^{(A)T}\mathcal{N}^{(A)T}_n, \nonumber\\
\chi^\mathrm{latt}_{0^\mp}&={(1\pm u)^2}\sum_{n=0}^{12} \bar{a}_{n}^{0^\mp}(1-u)^n \Delta_n^{0^\mp}\mathcal{N}^{0^\mp}_n.
\end{align}
where as well as including constant terms, $\bar{a}_{n}^{\delta}$ also allows for scale dependence through $\alpha_s$, as well as condensate contributions,
\begin{align}\label{ans}
\bar{a}_n^{\delta}=a_n^\delta\times\left(1+\beta_n^\delta {\alpha_s(\overline{m}_h)}+\kappa_n^\delta \frac{\langle \frac{\alpha_s}{\pi}G^2\rangle}{\overline{m}_h^3 \overline{m}_c}  \right)
\end{align}
where the $\overline{m}_h^3 \overline{m}_c$ factor was chosen to interpolate the expected quark mass dependence in both the $m_h\to m_b$ and $m_h\to m_c$ limits. We take $\langle \frac{\alpha_s}{\pi}G^2\rangle=0.02\mathrm{GeV}^4$ and use Gaussian priors of $0(2)$ for $\beta_n^\delta$ and $\kappa_n^\delta$.

$\Delta_n^\delta$ parameterises discretisation effects as
\begin{align}\label{deltandef}
\Delta_n^\delta=1+&\sum_{j=1}^6 b^{\delta}_{j,n}\left({am_h}\right)^{2j} +\sum_{j=1}^6c^{\delta}_{j,n}\left({am_c}\right)^{2j}\nonumber\\
+&\tilde{b}^{\delta}_{n}\left({am_h}\right)^{2} \mathrm{log}(am_h) +\tilde{c}^{\delta}_{n}\left({am_c}\right)^{2} \mathrm{log}(am_c).
\end{align}
We include terms accounting for log-enhanced discretisation effects which, due to the tree-level improvement of the HISQ action, are expected to enter at $\mathcal{O}(\alpha_s)$~\cite{Sommer:2022wac}. We take Gaussian priors of $0(2)$ for $\tilde{b}^{\delta}_{n}$ and $\tilde{c}^{\delta}_{n}$.

Because our simulation is done using staggered quarks, the correlation functions contain a time-oscillating contribution from time-doubled states with opposite parity~\cite{PhysRevD.75.054502}, with
\begin{align}\label{2ptsumexp}
C_\delta^\mathrm{latt}(t)=\sum_n |\lambda_n^\delta|^2 e^{-E^\delta_nt} - (-1)^t|\lambda^{\delta,\mathrm{osc}}_n|^2 e^{-E_n^{\delta,\mathrm{osc}}t}.
\end{align}
When we perform the sums over $t$ in~\cref{lattmoments}, the oscillating state contribution gives zero up to discretisation effects. For the $J^P=0^-$, $1^-$ and tensor currents, we expect the oscillating states to have $E_n^\mathrm{osc}>E_n$. In this case, the discretisation effects due to the oscillating state contribution are highly suppressed relative to the non-oscillating ground state. For the $J^P=0^+$, $1^+$ and axial-tensor currents, however, we expect $E_n^\mathrm{osc}<E_n$. In this case we can use the ground state parameters $\lambda_0^{(\mathrm{osc})}$ and $E_0^{(\mathrm{osc})}$, extracted from fits to~\cref{2ptsumexp} using the python package \textbf{corrfitter}~\cite{corrfitter}, to estimate the size of the discretisation effects from the oscillating state contribution to~\cref{lattmoments} relative to the nonoscillating ground state contribution. We find that this discretisation effect is expected to be largest on Sets 1 and 5, at the level of approximately $-15\%$, $-5\%$ and $-10\%$ for $\chi_{1^+}^\mathrm{latt}$, $\chi_{0^+}^\mathrm{latt}$ and $\chi_{AT}^\mathrm{latt}$ respectively. We therefore use a power series in $am_{q=h,c}$, as opposed to the more usual $am_q/\pi$, to capture these large discretisation effects, including up to $(am_q)^{12}$. For $b^{\delta}_{j,n}$ and $c^{\delta}_{j,n}$, we use Gaussian priors of $0(2)$.

In~\cite{Hoff:2011ge} it is observed that the expansion up to $\mathcal{O}((1-u)^{8})$ is indistinguishable from the full expressions for the leading order terms $\bar{C}_1^{(0),\delta}(u)$ from $u=0.8$ down to $u=u_\mathrm{phys}$. The expansion up to $\mathcal{O}((1-u)^{9})$ is also seen to reproduce the NLO and NNLO results well across the range $0.3\leq u\leq 0.8$ with deviations of $\approx 10\%$ close to $u=u_\mathrm{phys}$. Motivated by these observations, we include up to $(1-u)^{12}$ in our fit function. We have confirmed that this fit function reproduces the perturbative continuum results of~\cite{Grigo:2012ji} to $1$ part in $10^6$ across the range $u_\mathrm{phys}\leq u \leq 0.8$, with all $|a_{n}^{\delta}|< 0.01$. As such we use conservative Gaussian priors of $0.0(0.05)$ for each $a_{n}^{\delta}$ for terms with $n\leq 8$ and $0.0(0.025)$ for terms with $n> 8$, reflecting that these terms are only needed to capture the NLO and NNLO $u$-dependence of the perturbative results.
$\mathcal{N}^\delta$ accounts for valence and quark mass mistuning effects.
\begin{align}\label{Nndef}
\mathcal{N}^{\delta=1^{\pm},(A)T}_n =& \left(1 + A^{\delta} \delta_{m_c}^\mathrm{val}\right)\nonumber\\
&\times\left(1 + B^{\delta}_n \delta_{m_c}^{\mathrm{sea}}+ C^{\delta}_n \delta_{m_s}^\mathrm{sea}+ D^{\delta}_n \delta_{m_l}^\mathrm{sea}\right)\nonumber\\
\mathcal{N}^{\delta=0^{\pm}}_n =& \left(1 + B^{\delta}_n \delta_{m_c}^{\mathrm{sea}}+ C^{\delta}_n \delta_{m_s}^\mathrm{sea}+ D^{\delta}_n \delta_{m_l}^\mathrm{sea}\right)
\end{align}
with
\begin{align}
\delta_{m_c}^\mathrm{val} &= (am_c^\mathrm{val}-am_c^\mathrm{tuned})/am_c^\mathrm{tuned},\nonumber\\
\delta_{m_c}^{\mathrm{sea}} &= (am_{c}^\mathrm{sea}-am_c^\mathrm{val})/am_c^\mathrm{val},\nonumber\\
\delta_{m_{s}}^\mathrm{sea} &= (am_{s}^\mathrm{sea} - am_{s}^\mathrm{tuned})/(10am_{s}^\mathrm{tuned}),\nonumber\\
\delta_{m_{l}}^\mathrm{sea} &= (am_l^\mathrm{sea} - am_{s}^\mathrm{tuned}/[m_s/m_{l}]^\mathrm{phys})/(10am_{s}^\mathrm{tuned}),\label{deltatermseq}
\end{align}
and with $[m_s/m_{l}]^\mathrm{phys}=27.18(10)$ from~\cite{Bazavov:2017lyh}. When $m_c^\mathrm{val}=m_c^\mathrm{sea}$ the perturbative expressions for the susceptibilities are functions of only $u$, $\overline{m}_h$ and $\alpha_s(\overline{m}_h)$. Charm quark mistuning effects thus enter our calculation through the determination of $\overline{m}_h(\overline{m}_h)$ using the physical value of $\overline{m}_c(3\mathrm{GeV})$, as well as indirectly through the scale $\mu=\overline{m}_h$. The valence charm masses used here are well tuned, and the effect of the small mistuning on $\overline{m}_h(\overline{m}_h)$ leads to a negligible change in $\alpha_s(\overline{m}_h)$. Since the nonperturbative condensate contributions are expected to be small relative to the perturbative expressions, we also neglect their variation with the small valence charm mass mistunings. The only remaining place where mistuning effects may have a significant effect is the overall $1/\overline{m}_h^2$ appearing for the cases $\delta=1^\pm,(A)T$. For these cases, we take $\mathcal{N}^{\delta=1^\pm,(A)T}_n$ to contain only a single overall $\delta_{m_c}^\mathrm{val}$ factor. The relevant sea charm quark mistuning, which we denote $\delta_{m_c}^{\mathrm{sea}}$, is then the mistuning of the sea charm quark mass from the valence mass $m_c^\mathrm{val}$.

The tuned values of the quark masses are given by
\begin{equation}
am_c^\mathrm{tuned} = am_c^\mathrm{val}\frac{M_{\eta_c}^\mathrm{phys,QCD}}{M_{\eta_c}},
\end{equation}
where we use the pure QCD result $M_{\eta_c}^\mathrm{phys,QCD}=2.9783(11)\mathrm{GeV}$, computed using the results of~\cite{Hatton:2020qhk} for the $J/\psi$ hyperfine splitting in pure QCD and neglecting disconnected diagrams as we do here. To determine $M_{\eta_c}$, we generate $\eta_c$ correlation functions using local $\gamma^5\otimes\gamma^5$ spin-taste operators, using the valence charm masses given in~\cref{tab:valmasses}. We fit these correlation functions to
\begin{equation}
C_{\eta_c}^\mathrm{latt}(t)=\sum_{n=0}^{N_\mathrm{exp}=8} |\lambda_n|^2 e^{-M_nt},
\end{equation}
taking heuristic Gaussian priors of $0(0.75)$ for $\lambda_{n>0}$, $0.25(0.125)$ for $\lambda_0$, $0.75(0.6)\mathrm{GeV}$ for $M_{n+1}-M_n$ and $3.0(0.75)\mathrm{GeV}$ for $M_0=M_{\eta_c}$. The values of $M_{\eta_c}$ resulting from this fit are given in lattice units in~\cref{tab:Metactable}, where we see excellent agreement with the values determined in~\cite{Hatton:2020qhk}, allowing for the small differences in valence masses on sets 1 and 4.

\begin{table}
\centering
\caption{$\eta_c$ masses in lattice units, used to determine $am_c^\mathrm{tuned}$. \label{tab:Metactable}}
\begin{tabular}{c c }\hline
 Set & $aM_{\eta_c}$  \\ \hline
1 & $1.364965(66)$     \\\hline
2 & $0.896644(80)$     \\\hline
3 & $0.666886(75)$     \\\hline
4 & $0.49423(16)$     \\ \hline
5 & $1.33045(97)$     \\ \hline
\end{tabular}
\end{table} 

We take
\begin{equation}
am_s^\mathrm{tuned} = am_s^\mathrm{val}\left(\frac{M_{\eta_s}^\mathrm{phys}}{M_{\eta_s}}\right)^2,
\end{equation}
where we use the values of $M_{\eta_s}$ given in~\cite{EuanBsDsstar}. Since these values are very precise, and since we expect sea quark mass mistuning effects to be small, we neglect their correlations with our other data. We take priors of $0(2)$ for each $A^{\delta=1^\pm,(A)T}$, and $0.0(0.5)$ for $C_n$ and $D_n$ to reflect the fact that the corresponding sea quark mistuning effects appear at next-to-leading order in $\alpha_s$. We take a prior of $0(0.1)$ for $B_n$, to reflect the results of the analysis of sea charm quark mistuning effects on $w_0$ in~\cite{PhysRevD.91.054508}.

\section{Results}
\begin{figure}
\centering
\includegraphics[scale=0.18]{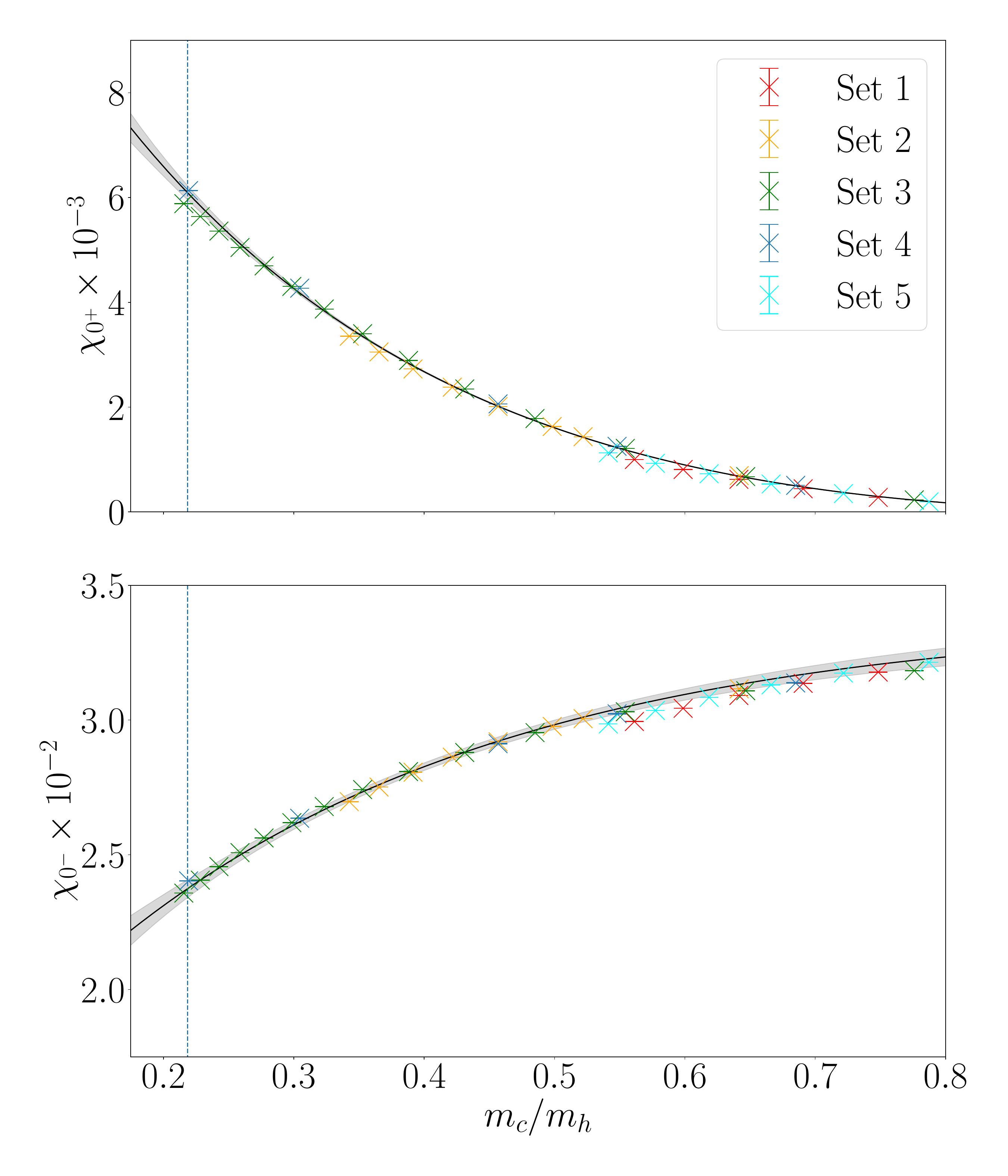}
\caption{\label{resplots0}Plot showing our lattice data points for the (pseudo)scalar susceptibilities, together with the result of our chiral continuum extrapolation (grey band) as a function of $m_c/m_h$ for the (pseudo)scalar susceptibilities.}
\end{figure}
\begin{figure}
\centering
\includegraphics[scale=0.18]{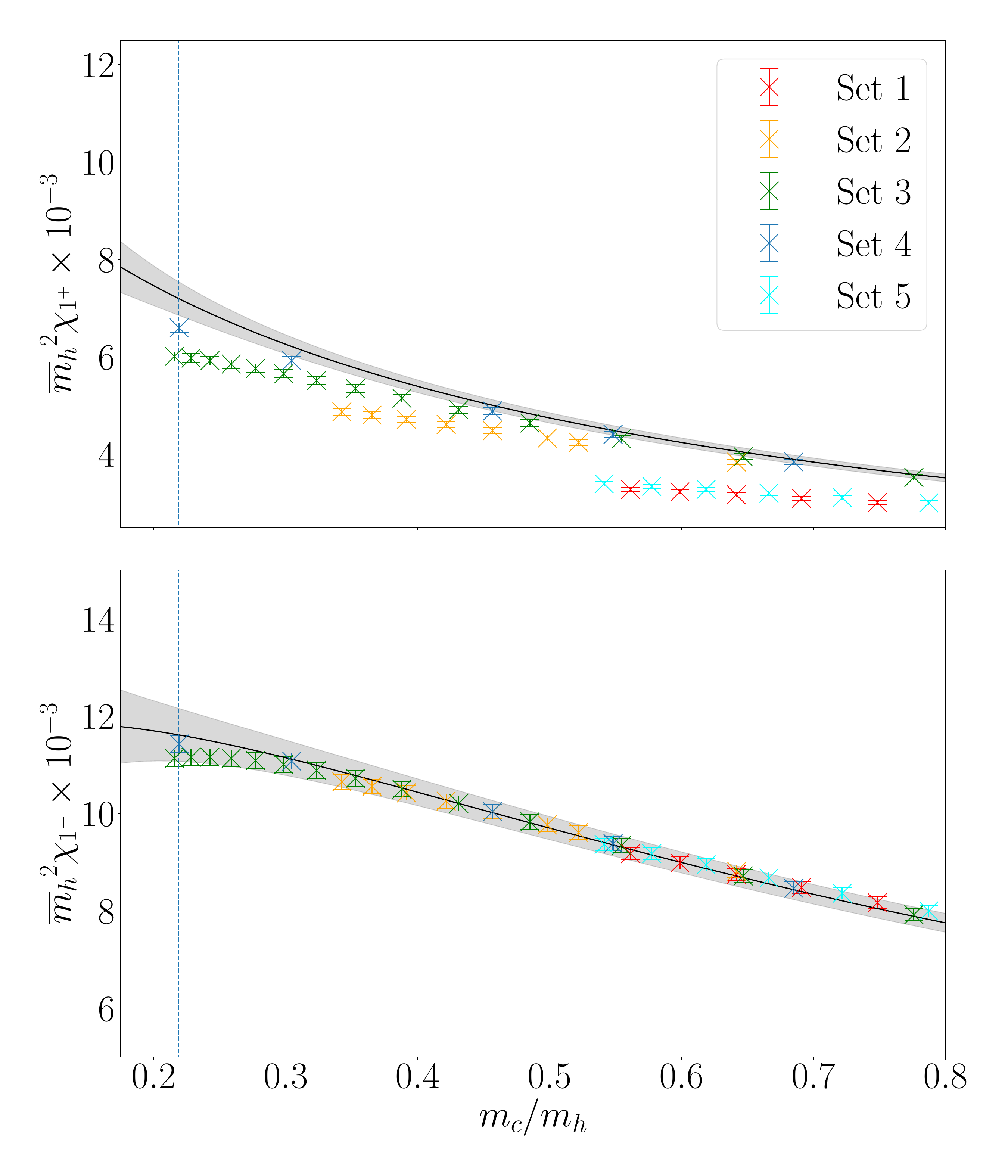}
\caption{\label{resplots1}Plot showing our lattice data points for the (axial-)vector susceptibilities, together with the result of our chiral continuum extrapolation (grey band) as a function of $m_c/m_h$ for the (axial-)vector susceptibilities.}
\end{figure}

\begin{figure}
\centering
\includegraphics[scale=0.18]{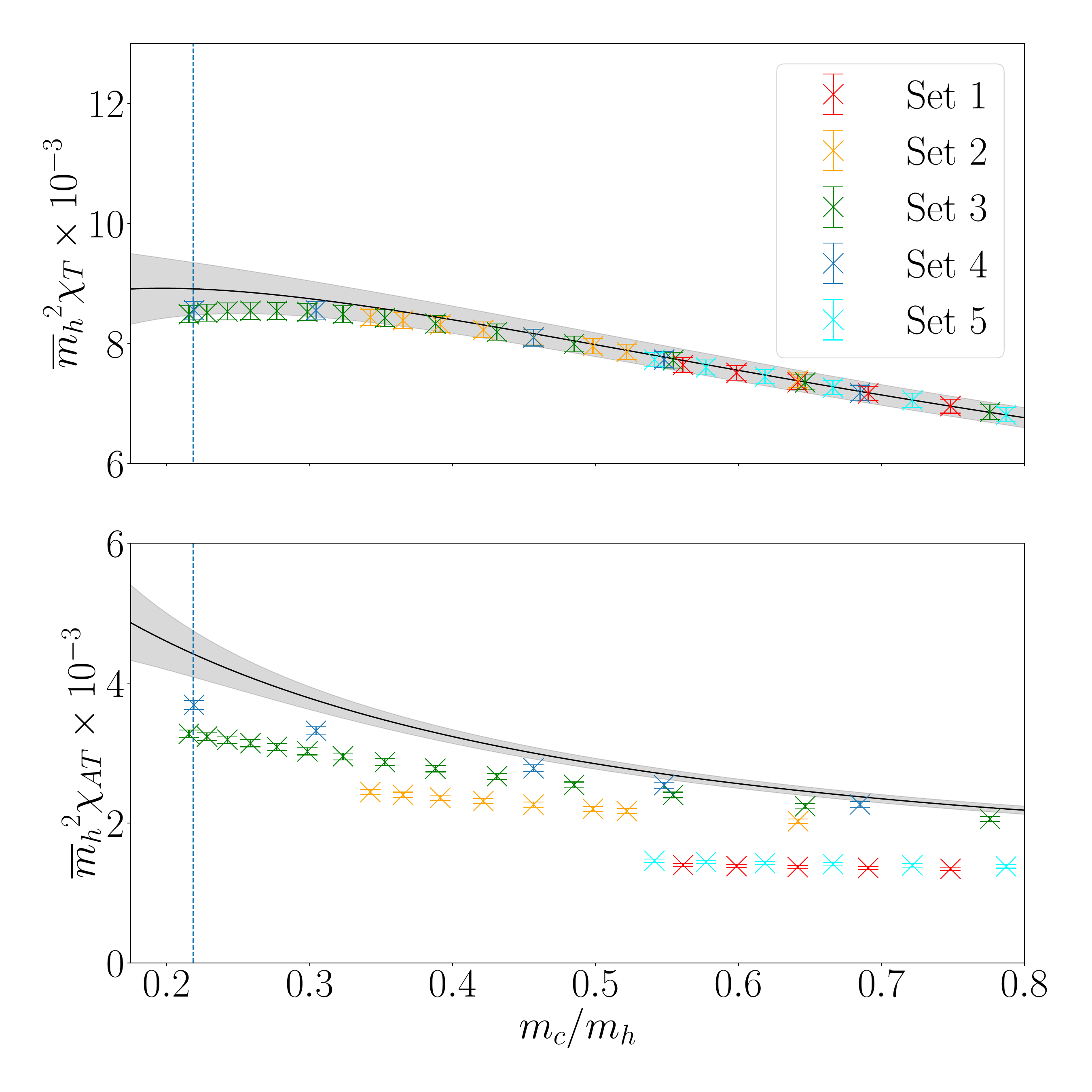}
\caption{\label{tensors}Plot showing our lattice data points together with the result of our chiral continuum extrapolation (grey band) as a function of $m_c/m_h$ for the (axial-)tensor susceptibilities. Note that the discretisation effects appearing in the axial-tensor susceptibility are somewhat larger than expected from the estimate using ground state parameters extracted from correlator fits.}
\end{figure}

We use the python package \textbf{lsqfit}~\cite{lsqfit} to perform the fit to~\cref{continuumfitfunc}. Our lattice data points and continuum extrapolated susceptibilities for the (pseudo)scalar and (axial-)vector susceptibilities are plotted in~\cref{resplots0,resplots1}.
Our lattice data points for the tensor susceptibilities are shown in~\cref{tensors}, together with the result of our chiral continuum extrapolation. The fit has $\chi^2/\mathrm{dof}=0.89$, which we estimate using svd and prior noise~\cite{Dowdall:2019bea}, and a corresponding $Q$-value of $Q = 0.89$. We see that the discretisation effects are visibly larger for $\chi_{1^+}$ as expected~(see~\cref{sec:contextrap}). 

We find, for the physical $b$-quark,
\begin{align}\label{chires}
\overline{m}_b^2 \times\chi_{1^+}&=\schionep,\nonumber\\
\overline{m}_b^2 \times\chi_{1^-}&=\schionem,\nonumber\\
\chi_{0^-}&=\schizm,\nonumber\\
\chi_{0^+}&=\schizp,
\end{align}
and for the tensor susceptibilities,
\begin{align}\label{chiresT}
\overline{m}_b^2 \times\chi_{T}&=\schiT,\nonumber\\
\overline{m}_b^2 \times\chi_{AT}&=\schiAT.
\end{align}

For ease of comparison to other results, we also give the (axial-)vector and (axial-)tensor susceptibilities with the factor of $1/\overline{m}_b^2$ included. We find

\begin{align}\label{chiresMh}
\chi_{1^+}&=\schionepMh,\nonumber\\
\chi_{1^-}&=\schionemMh,\nonumber\\
\chi_{T}&=\schiTMh,\nonumber\\
\chi_{AT}&=\schiATMh.
\end{align}

In order to provide self-contained results, we generate synthetic data across the full range of $u$ between 0.8 and $m_c/m_b$ and fit this data using a simple power series in $1-u$ up to $(1-u)^{12}$, as in~\cref{continuumfitfunc}, without any factors of $\alpha_s$ and $\langle \frac{\alpha_s}{\pi}G^2\rangle$. We find that the susceptibilities computed from the results of this fit are indistinguishable from our full results, and we provide the posterior distributions for the coefficients in the file \textbf{susceptibilities\_u12.pydat} in the supplementary material, as well as the python script \textbf{load\_chi\_u12.py}, which loads the correlated parameters from \textbf{susceptibilities\_u12.pydat} and computes the continuum susceptibilities.

\subsection{Tests of the Stability of the Analysis}
\label{stabsec}
In order to demonstrate the robustness of our results to changes in the chiral continuum fit function~\cref{continuumfitfunc}, we repeat the above analysis for several variations of the fit function. We show results obtaining using fits with higher orders of $1-u$ included, with higher orders in $am_{q=h,c}$ included, as well as a fit including only up to $(1-u)^8$ and $(am_{q=h,c})^8$. In addition to these variations, we also show the results of fits excluding the $a^2\mathrm{log}(a)$ terms from~\cref{deltandef}, as well as excluding the terms proportional to $\alpha_s(\bar{m}_h)$ and $\langle \frac{\alpha_s}{\pi}G^2\rangle$ in~\cref{ans}. The results of these fits are shown for $u=0.2184$, $u=0.5$ and $u=0.8$ in~\cref{stability1m1p,stability0m0p,stabilityTAT} in~\cref{sec:appstabplots}, where we see that our results vary only very slightly at each point for each different chiral continuum fitting strategy.

\subsection{Comparison to Existing Results}

The $\overline{b}c$ susceptibilities are expected to receive only extremely small nonperturbative condensate corrections, at the level of $\approx 0.05\%$ for the physical $b$-quark mass~\cite{Boyd:1997kz}. As such, we expect that there should be good agreement between our lattice results for the (pseudo)scalar and (axial-)vector susceptibilities and those determined using the results of~\cite{Grigo:2012ji}.
\begin{figure}
\centering
\includegraphics[scale=0.18]{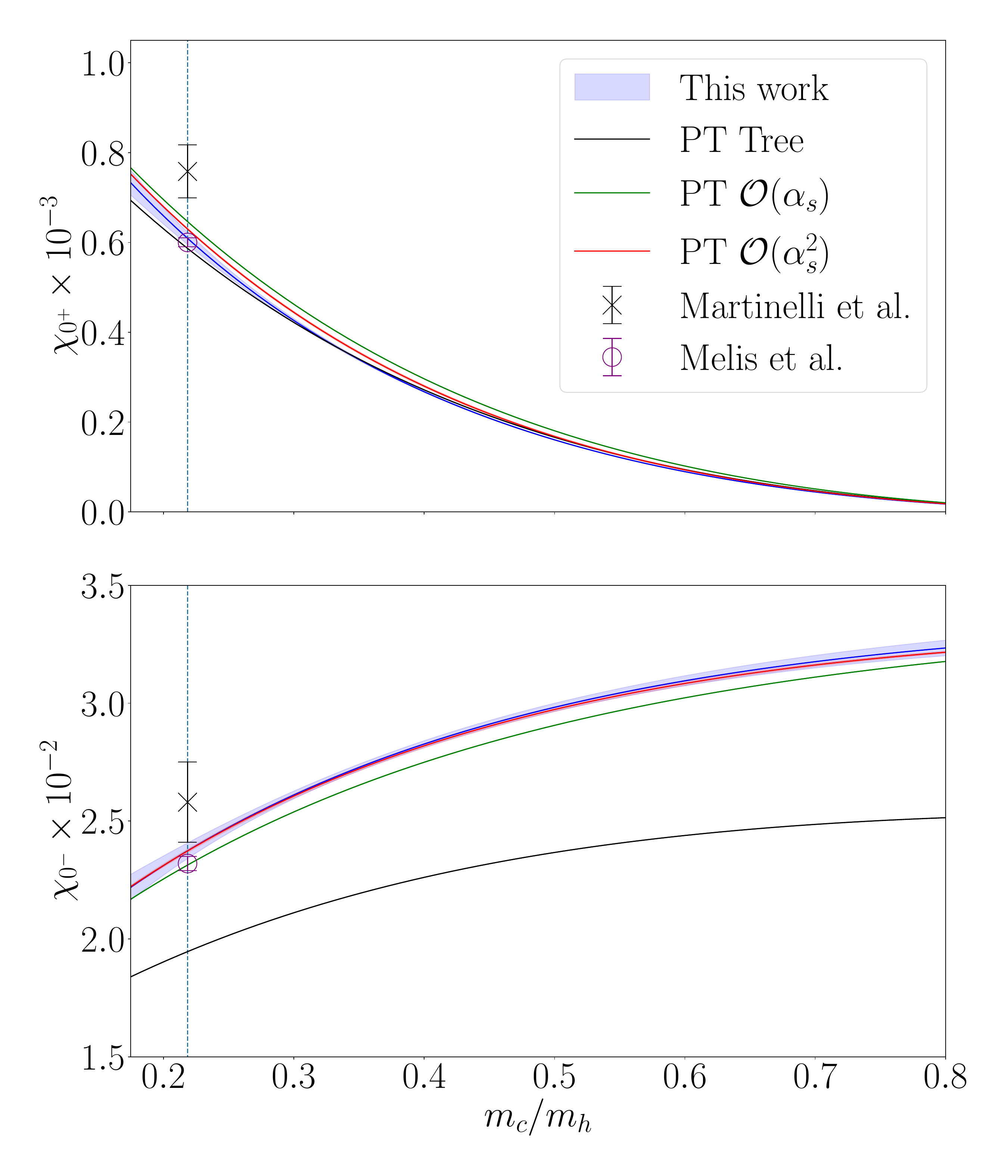}
\caption{\label{oursromepert0} Our chiral continuum fit results for the (pseudo)scalar susceptibilities (blue band) compared to the perturbative result at tree-level~(black line), $\mathcal{O}(\alpha_s)$~(green line) and $\mathcal{O}(\alpha_s^2)$~(red line). We add the leading condensate contribution from~\cite{Boyd:1997kz} to the $\mathcal{O}(\alpha_s^2)$ result in red, though this has a very small effect. The red band showing the uncertainty on the $\mathcal{O}(\alpha_s^2)$ result is equal to $(\alpha_s/\pi)^3$ multipled by the root-mean-square of the three known coefficients. We also include the results of~\cite{Martinelli:2021frl} and~\cite{Melis:2024wpb} for comparison. We see that our results are very close to the perturbation theory across the full range of $m_c/m_h$ considered.}
\end{figure}

\begin{figure}
\centering
\includegraphics[scale=0.18]{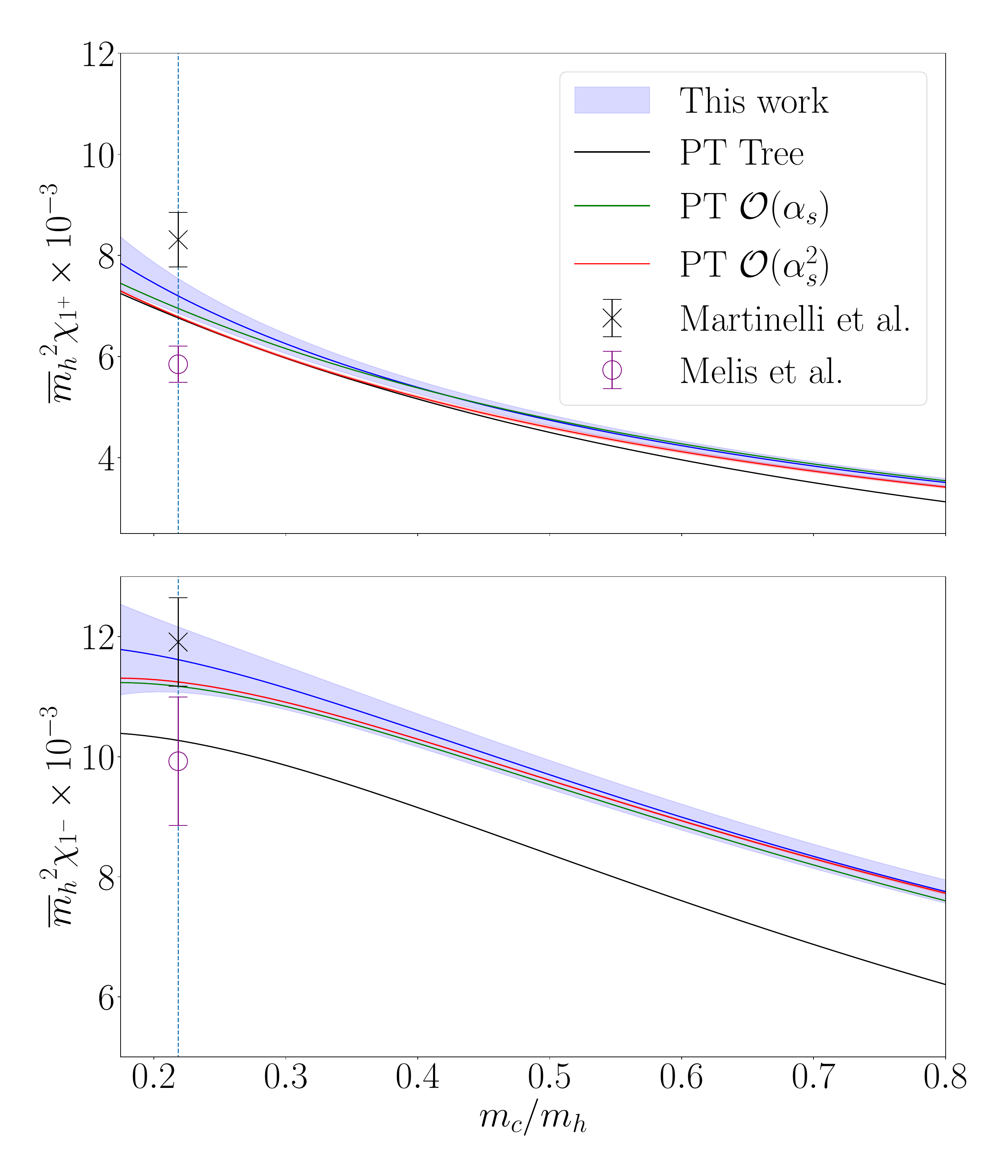}
\caption{\label{oursromepert1} Our chiral continuum fit results for the (axial-) vector susceptibilities (blue band) compared to the perturbative result at tree-level~(black line), $\mathcal{O}(\alpha_s)$~(green line) and $\mathcal{O}(\alpha_s^2)$~(red line). We add the leading condensate contribution from~\cite{Boyd:1997kz} to the $\mathcal{O}(\alpha_s^2)$ result in red, though this has a very small effect. The red band showing the uncertainty on the $\mathcal{O}(\alpha_s^2)$ result is equal to  $(\alpha_s/\pi)^3$ multipled by the root-mean-square of the three known coefficients. We also include the results of~\cite{Martinelli:2021frl} and~\cite{Melis:2024wpb} for comparison. We see that our results are very close to the perturbation theory across the full range of $m_c/m_h$ considered.}
\end{figure}

\begin{figure}
\centering
\includegraphics[scale=0.18]{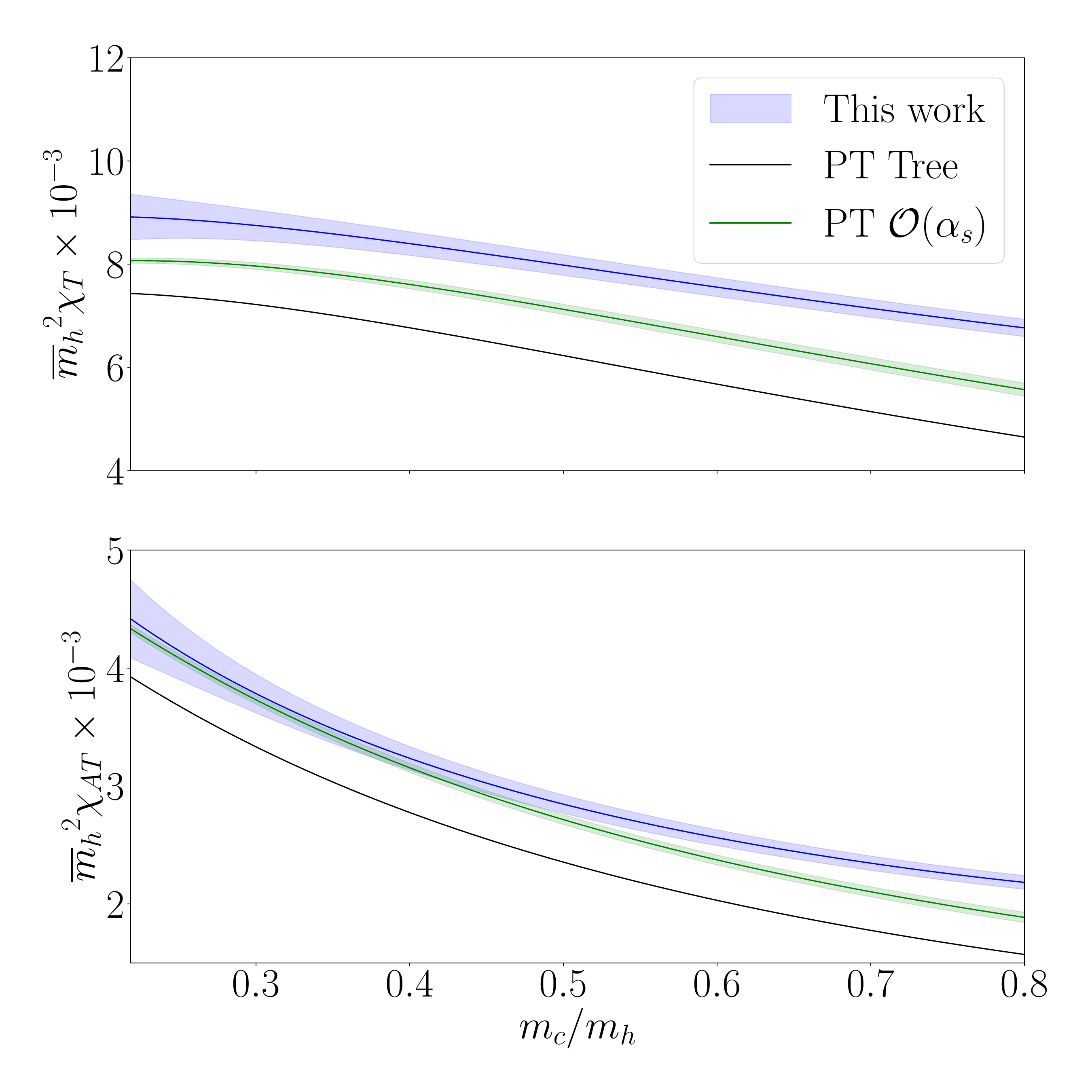}
\caption{\label{oursromepertT} Our chiral continuum fit results for the (axial-)tensor susceptibilities (blue band) compared to the perturbative result at tree-level~(black line) and $\mathcal{O}(\alpha_s)$~(green line) computed in~\cite{Bharucha:2010im}. The green band showing the uncertainty on the $\mathcal{O}(\alpha_s)$ result is equal to  $\alpha_s^2$ multipled by the root-mean-square of the two known coefficients. We see that our results are in reasonable agreement for the axial-tensor susceptibility, but disagree for the tensor susceptibility.}
\end{figure}

Our continuum results for the (axial-)vector and (pseudo)scalar susceptibilities are plotted in~\cref{oursromepert0,oursromepert1}, together with the LO, NLO and NNLO results determined using the results of~\cite{Grigo:2012ji} that we describe in~\cref{sec:theory}. In addition to the NNLO perturbative result, we include the leading order condensate contribution given in~\cite{Boyd:1997kz}. To evaluate these expressions, which are given in terms of the pole masses, we use the two-loop matching between the $\overline{\mathrm{MS}}$ and pole masses from~\cite{Gray:1990yh}, allowing a $10\%$ uncertainty for renormalon effects. We see that our lattice results, plotted as the blue band, are very close to the result including NNLO perturbation theory and leading condensate terms across the full range of $u$ values considered. Taking each susceptibility in isolation, we find reasonable agreement between our results and the perturbation theory for the vector, axial-vector and pseudoscalar cases across the full range of $u$. Our result for the scalar susceptibility is in slight disagreement with the perturbative result in the region where $u\approx 0.3$. 

The lattice results from~\cite{Martinelli:2021frl} and~\cite{Melis:2024wpb} are also plotted in~\cref{oursromepert0,oursromepert1}. We see good agreement between our results and those of~\cite{Martinelli:2021frl} for $\chi_{1^-}$, but disagreement at the level of $1-2\sigma$ for $\chi_{1^+}$, $\chi_{0^-}$ and $\chi_{0^+}$. For the more recent results of~\cite{Melis:2024wpb}, we see excellent agreement for $\chi_{0^-}$ and $\chi_{0^+}$, mild tension at the level of $1\sigma$ for $\chi_{1^-}$, but poor agreement for $\chi_{1^+}$.

For the (axial-)tensor cases, the susceptibilities have been computed perturbatively to $\mathcal{O}(\alpha_s)$~\cite{Bharucha:2010im}. We plot our continuum results for the (axial-)tensor together with the perturbative results in~\cref{oursromepertT}. We see good agreement between our results and the perturbation theory for the axial-tensor susceptibility for $m_h\to m_b$, but poor agreement for $m_h\to m_c$. For the tensor susceptibility, we find significant disagreement with the NLO perturbative result across the full range of $u$.

\section{Conclusion}
We have computed the full set of (pseudo)scalar, (axial-)vector and (axial-)tensor susceptibilities, $\chi_{1^-}$, $\chi_{1^+}$, $\chi_{0^-}$, $\chi_{0^+}$, $\chi_{T}$ and $\chi_{AT}$, between $u=u_\mathrm{phys}$ and $u=0.8$, using the heavy-HISQ method, including up/down quarks, and physically tuned strange and charm quarks in the sea. Importantly, we include here a gauge field ensemble with $a\approx 0.03\mathrm{fm}$, sufficiently small for the physical $b$-quark mass to be reached, with $am_b\approx 0.625$. 

We find that our results for the pseudoscalar and (axial-)vector susceptibilities are in agreement with the 3-loop perturbation theory results~\cite{Grigo:2012ji}, while the scalar susceptibility exhibits some tension. Our results demonstrate the reliability of this method of computing susceptibilities on the lattice. We find that the tensor and axial-tensor susceptibilities at the physical $b$-quark mass are roughly $1/3$ smaller than the vector and axial-vector susceptibilities respectively. This is to be expected from the similar size difference seen in the OPE results for the $\bar{b}s$ tensor and axial tensor in~\cite{Bharucha:2010im} together with the observation that the fourth moments of the $\bar{h}s$ and $\bar{h}c$ correlators computed in~\cite{Davies:2018hmw} differ by only a few percent for the largest values of $am_h$. We find reasonable agreement with the NLO perturbation theory for the axial-tensor susceptibility, but for the tensor our results are in disagreement with the perturbative result, as seen in~\cref{oursromepertT}.

The results of this work will allow future lattice calculations of $b\to c$ form factors, for both mesonic and baryonic decays, to use dispersively bounded parameterisations for all form factors, for varying heavy quark mass between $1.25 \times m_c$ and $m_b$, using lattice results for all inputs. This work will also lead to future lattice calculations of less well-known quantities entering the dispersive bounds for other hadronic form factors, such as those needed for $b\to s$ decays~\cite{Gubernari:2023puw} where perturbative calculations of the susceptibilities are less reliable due to the much more sizeable condensate contributions.

\subsection*{\bf{Acknowledgements}}

We are grateful to the MILC Collaboration for the use
of their configurations and code. We thank C. T.H. Davies, C. Bouchard,
B. Colquhoun, D. van Dyk, M. Jung and M. Bordone for useful discussions. Computing was done on
the Cambridge service for Data Driven Discovery (CSD3),
part of which is operated by the University of Cambridge
Research Computing on behalf of the DIRAC HPC Facility
of the Science and Technology Facilities Council (STFC).
The DIRAC component of CSD3 was funded by BEIS
capital funding via STFC capital Grants No. ST/P002307/1
and No. ST/R002452/1 and by STFC operations Grant
No. ST/R00689X/1. DiRAC is part of the national
e-infrastructure. We are grateful to the CSD3 support staff
for assistance. Funding for this work came from UK Science
and Technology Facilities Council Grants No. ST/L000466/1 and No. ST/P000746/1 and Engineering and Physical
Sciences Research Council Project No. EP/W005395/1.

\begin{appendix}
\section{Lattice Data}
\label{sec:applattdat}
Here we give our raw lattice results for the susceptibilities $\chi_{\delta}^\mathrm{latt}(q^2=0)$ on each ensemble. Results for the (pseudo)scalar and (axial-)vector susceptibilities are given in~\cref{tab:chilattres1,tab:chilattres2,tab:chilattres3,tab:chilattres4,tab:chilattres5}, while those for the (axial-)tensor currents are given in~\cref{tab:Tchilattres1,tab:Tchilattres2,tab:Tchilattres3,tab:Tchilattres4,tab:Tchilattres5}.

\begin{table}
\caption{Susceptibilities $\chi_\delta$, defined in~\cref{lattmoments}, for each value of $am_h$ used on set 1.\label{tab:chilattres1}}
\begin{tabular}{c c c c c }\hline
 $am_h$ &$\chi_{1^+}\cdot 10^{4}\mathrm{GeV}^2$ & $\chi_{1^-}\cdot 10^{4}\mathrm{GeV}^2$ &$\chi_{0^-}\cdot 10^{2}$ &$\chi_{0^+}\cdot 10^{3}$   \\\hline 
0.55  & $ 28.99(44) $   & $ 78.0(1.2) $    &$ 3.2153(4) $ &$ 0.13762(2) $\\
0.6  & $ 30.02(44) $   & $ 81.7(1.2) $    &$ 3.1777(4) $ &$ 0.27640(3) $\\
0.65  & $ 30.90(45) $   & $ 84.8(1.2) $    &$ 3.1360(4) $ &$ 0.44133(4) $\\
0.7  & $ 31.63(45) $   & $ 87.5(1.2) $    &$ 3.0910(3) $ &$ 0.62176(6) $\\
0.75  & $ 32.23(45) $   & $ 89.8(1.3) $    &$ 3.0436(3) $ &$ 0.80977(7) $\\
0.8  & $ 32.72(45) $   & $ 91.7(1.3) $    &$ 2.9944(3) $ &$ 0.99950(8) $\\
\end{tabular}
\end{table}
\begin{table}
\caption{Susceptibilities $\chi_\delta$, defined in~\cref{lattmoments}, for each value of $am_h$ used on set 2.\label{tab:chilattres2}}
\begin{tabular}{c c c c c }\hline
 $am_h$ &$\chi_{1^+}\cdot 10^{4}\mathrm{GeV}^2$ & $\chi_{1^-}\cdot 10^{4}\mathrm{GeV}^2$ &$\chi_{0^-}\cdot 10^{2}$ &$\chi_{0^+}\cdot 10^{3}$   \\\hline 
0.427  & $ 38.34(56) $   & $ 88.1(1.3) $    &$ 3.1158(5) $ &$ 0.69297(9) $\\
0.525  & $ 42.42(61) $   & $ 96.1(1.4) $    &$ 3.0050(4) $ &$ 1.4349(1) $\\
0.55  & $ 43.28(62) $   & $ 97.7(1.4) $    &$ 2.9763(4) $ &$ 1.6290(2) $\\
0.6  & $ 44.80(63) $   & $ 100.4(1.4) $    &$ 2.9190(4) $ &$ 2.0113(2) $\\
0.65  & $ 46.08(65) $   & $ 102.5(1.4) $    &$ 2.8623(4) $ &$ 2.3793(2) $\\
0.7  & $ 47.13(66) $   & $ 104.2(1.5) $    &$ 2.8063(3) $ &$ 2.7275(2) $\\
0.75  & $ 47.99(68) $   & $ 105.5(1.5) $    &$ 2.7512(3) $ &$ 3.0527(2) $\\
0.8  & $ 48.67(69) $   & $ 106.5(1.5) $    &$ 2.6971(3) $ &$ 3.3532(2) $\\
\end{tabular}
\end{table}
\begin{table}
\caption{Susceptibilities $\chi_\delta$, defined in~\cref{lattmoments}, for each value of $am_h$ used on set 3.\label{tab:chilattres3}}
\begin{tabular}{c c c c c }\hline
 $am_h$ &$\chi_{1^+}\cdot 10^{4}\mathrm{GeV}^2$ & $\chi_{1^-}\cdot 10^{4}\mathrm{GeV}^2$ &$\chi_{0^-}\cdot 10^{2}$ &$\chi_{0^+}\cdot 10^{3}$   \\\hline 
0.25  & $ 35.19(57) $   & $ 79.2(1.3) $    &$ 3.1833(8) $ &$ 0.23015(5) $\\
0.3  & $ 39.45(61) $   & $ 87.2(1.4) $    &$ 3.1088(7) $ &$ 0.6733(1) $\\
0.35  & $ 43.16(66) $   & $ 93.4(1.4) $    &$ 3.0306(6) $ &$ 1.2119(2) $\\
0.4  & $ 46.36(70) $   & $ 98.3(1.5) $    &$ 2.9534(5) $ &$ 1.7818(2) $\\
0.45  & $ 49.12(74) $   & $ 102.1(1.5) $    &$ 2.8792(4) $ &$ 2.3478(3) $\\
0.5  & $ 51.46(77) $   & $ 105.0(1.6) $    &$ 2.8087(4) $ &$ 2.8909(3) $\\
0.55  & $ 53.45(80) $   & $ 107.2(1.6) $    &$ 2.7420(4) $ &$ 3.4008(3) $\\
0.6  & $ 55.11(83) $   & $ 108.9(1.6) $    &$ 2.6789(3) $ &$ 3.8724(3) $\\
0.65  & $ 56.48(85) $   & $ 110.1(1.7) $    &$ 2.6192(3) $ &$ 4.3037(3) $\\
0.7  & $ 57.60(87) $   & $ 110.9(1.7) $    &$ 2.5623(3) $ &$ 4.6945(3) $\\
0.75  & $ 58.49(88) $   & $ 111.4(1.7) $    &$ 2.5080(2) $ &$ 5.0460(3) $\\
0.8  & $ 59.18(90) $   & $ 111.6(1.7) $    &$ 2.4560(2) $ &$ 5.3597(3) $\\
0.85  & $ 59.70(91) $   & $ 111.5(1.7) $    &$ 2.4060(2) $ &$ 5.6379(3) $\\
0.9  & $ 60.06(92) $   & $ 111.3(1.7) $    &$ 2.3578(2) $ &$ 5.8829(3) $\\
\end{tabular}
\end{table}
\begin{table}
\caption{Susceptibilities $\chi_\delta$, defined in~\cref{lattmoments}, for each value of $am_h$ used on set 4.\label{tab:chilattres4}}
\begin{tabular}{c c c c c }\hline
 $am_h$ &$\chi_{1^+}\cdot 10^{4}\mathrm{GeV}^2$ & $\chi_{1^-}\cdot 10^{4}\mathrm{GeV}^2$ &$\chi_{0^-}\cdot 10^{2}$ &$\chi_{0^+}\cdot 10^{3}$   \\\hline 
0.2  & $ 38.37(59) $   & $ 84.6(1.3) $    &$ 3.138(1) $ &$ 0.5065(1) $\\
0.25  & $ 44.04(66) $   & $ 93.8(1.4) $    &$ 3.0231(9) $ &$ 1.2515(3) $\\
0.3  & $ 48.84(72) $   & $ 100.4(1.5) $    &$ 2.9122(8) $ &$ 2.0614(4) $\\
0.45  & $ 59.18(88) $   & $ 110.8(1.6) $    &$ 2.6354(5) $ &$ 4.2698(6) $\\
0.625  & $ 65.92(99) $   & $ 114.3(1.7) $    &$ 2.4031(3) $ &$ 6.1319(5) $\\
\end{tabular}
\end{table}
\begin{table}
\caption{Susceptibilities $\chi_\delta$, defined in~\cref{lattmoments}, for each value of $am_h$ used on set 5.\label{tab:chilattres5}}
\begin{tabular}{c c c c c }\hline
 $am_h$ &$\chi_{1^+}\cdot 10^{4}\mathrm{GeV}^2$ & $\chi_{1^-}\cdot 10^{4}\mathrm{GeV}^2$ &$\chi_{0^-}\cdot 10^{2}$ &$\chi_{0^+}\cdot 10^{3}$   \\\hline 
0.55  & $ 29.97(44) $   & $ 80.0(1.2) $    &$ 3.2141(2) $ &$ 0.19143(1) $\\
0.6  & $ 31.06(45) $   & $ 83.6(1.2) $    &$ 3.1742(2) $ &$ 0.35010(2) $\\
0.65  & $ 31.98(45) $   & $ 86.8(1.2) $    &$ 3.1306(2) $ &$ 0.53221(3) $\\
0.7  & $ 32.74(46) $   & $ 89.5(1.2) $    &$ 3.0841(2) $ &$ 0.72738(3) $\\
0.75  & $ 33.38(46) $   & $ 91.7(1.3) $    &$ 3.0355(2) $ &$ 0.92792(4) $\\
0.8  & $ 33.89(46) $   & $ 93.6(1.3) $    &$ 2.9853(2) $ &$ 1.12822(5) $\\
\end{tabular}
\end{table}
\begin{table}
\caption{Tensor and axial-tensor susceptibilities $\chi_\delta$, defined in~\cref{lattmoments}, for each value of $am_h$ used on set 1.\label{tab:Tchilattres1}}
\begin{tabular}{c c c }\hline
 $am_h$ &$\chi_{T}\cdot 10^{4}\mathrm{GeV}^2$ & $\chi_{AT}\cdot 10^{4}\mathrm{GeV}^2$   \\\hline 
0.55  & $ 67.1(1.2) $   & $ 13.31(23) $\\
0.6  & $ 69.6(1.2) $   & $ 13.45(23) $\\
0.65  & $ 71.7(1.2) $   & $ 13.57(23) $\\
0.7  & $ 73.5(1.2) $   & $ 13.70(23) $\\
0.75  & $ 75.1(1.2) $   & $ 13.84(23) $\\
0.8  & $ 76.4(1.2) $   & $ 13.99(23) $\\
\end{tabular}
\end{table}
\begin{table}
\caption{Tensor and axial-tensor susceptibilities $\chi_\delta$, defined in~\cref{lattmoments}, for each value of $am_h$ used on set 2.\label{tab:Tchilattres2}}
\begin{tabular}{c c c }\hline
 $am_h$ &$\chi_{T}\cdot 10^{4}\mathrm{GeV}^2$ & $\chi_{AT}\cdot 10^{4}\mathrm{GeV}^2$   \\\hline 
0.427  & $ 73.9(1.2) $   & $ 20.24(34) $\\
0.525  & $ 78.6(1.3) $   & $ 21.70(35) $\\
0.55  & $ 79.5(1.3) $   & $ 22.02(36) $\\
0.6  & $ 81.1(1.3) $   & $ 22.61(36) $\\
0.65  & $ 82.3(1.3) $   & $ 23.13(37) $\\
0.7  & $ 83.2(1.3) $   & $ 23.60(38) $\\
0.75  & $ 83.9(1.3) $   & $ 24.03(39) $\\
0.8  & $ 84.4(1.4) $   & $ 24.42(39) $\\
\end{tabular}
\end{table}
\begin{table}
\caption{Tensor and axial-tensor susceptibilities $\chi_\delta$, defined in~\cref{lattmoments}, for each value of $am_h$ used on set 3.\label{tab:Tchilattres3}}
\begin{tabular}{c c c }\hline
 $am_h$ &$\chi_{T}\cdot 10^{4}\mathrm{GeV}^2$ & $\chi_{AT}\cdot 10^{4}\mathrm{GeV}^2$   \\\hline 
0.25  & $ 68.6(1.2) $   & $ 20.57(37) $\\
0.3  & $ 73.5(1.3) $   & $ 22.39(38) $\\
0.35  & $ 77.2(1.3) $   & $ 24.01(40) $\\
0.4  & $ 79.9(1.3) $   & $ 25.43(42) $\\
0.45  & $ 81.9(1.4) $   & $ 26.68(44) $\\
0.5  & $ 83.3(1.4) $   & $ 27.76(46) $\\
0.55  & $ 84.3(1.4) $   & $ 28.70(47) $\\
0.6  & $ 84.9(1.4) $   & $ 29.52(49) $\\
0.65  & $ 85.3(1.4) $   & $ 30.23(50) $\\
0.7  & $ 85.4(1.4) $   & $ 30.86(51) $\\
0.75  & $ 85.5(1.4) $   & $ 31.41(52) $\\
0.8  & $ 85.4(1.4) $   & $ 31.90(53) $\\
0.85  & $ 85.2(1.4) $   & $ 32.34(54) $\\
0.9  & $ 84.9(1.4) $   & $ 32.74(55) $\\
\end{tabular}
\end{table}
\begin{table}
\caption{Tensor and axial-tensor susceptibilities $\chi_\delta$, defined in~\cref{lattmoments}, for each value of $am_h$ used on set 4.\label{tab:Tchilattres4}}
\begin{tabular}{c c c }\hline
 $am_h$ &$\chi_{T}\cdot 10^{4}\mathrm{GeV}^2$ & $\chi_{AT}\cdot 10^{4}\mathrm{GeV}^2$   \\\hline 
0.2  & $ 71.8(1.3) $   & $ 22.64(41) $\\
0.25  & $ 77.3(1.4) $   & $ 25.40(45) $\\
0.3  & $ 81.0(1.4) $   & $ 27.81(48) $\\
0.45  & $ 85.5(1.5) $   & $ 33.17(57) $\\
0.625  & $ 85.6(1.5) $   & $ 36.87(64) $\\
\end{tabular}
\end{table}
\begin{table}
\caption{Tensor and axial-tensor susceptibilities $\chi_\delta$, defined in~\cref{lattmoments}, for each value of $am_h$ used on set 5.\label{tab:Tchilattres5}}
\begin{tabular}{c c c }\hline
 $am_h$ &$\chi_{T}\cdot 10^{4}\mathrm{GeV}^2$ & $\chi_{AT}\cdot 10^{4}\mathrm{GeV}^2$   \\\hline 
0.55  & $ 68.1(1.2) $   & $ 13.78(23) $\\
0.6  & $ 70.6(1.2) $   & $ 13.95(23) $\\
0.65  & $ 72.7(1.2) $   & $ 14.11(23) $\\
0.7  & $ 74.5(1.2) $   & $ 14.26(23) $\\
0.75  & $ 76.0(1.2) $   & $ 14.42(23) $\\
0.8  & $ 77.3(1.2) $   & $ 14.59(23) $\\
\end{tabular}
\end{table}

\section{Stability Plots}
\label{sec:appstabplots}
\begin{figure*}
\centering
\includegraphics[scale=1]{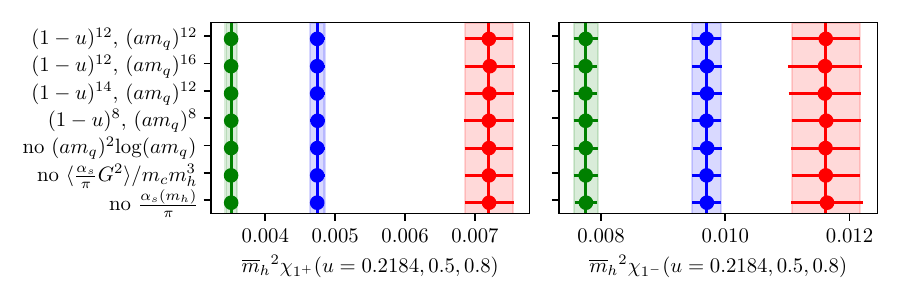}
\caption{\label{stability1m1p} Comparison of $\chi_{1^-}(u)$(left) and $\chi_{1^+}(u)$(right) at $u=0.2184$~(red), $u=0.5$~(blue) and $u=0.8$~(green) computed using the variations of the fit described in~\cref{stabsec}, indicated on the vertical axis. The topmost value and filled band correspond to our final results. We see that our results vary only very slightly for these different methods of performing the extrapolation.}
\end{figure*}

\begin{figure*}
\centering
\includegraphics[scale=1]{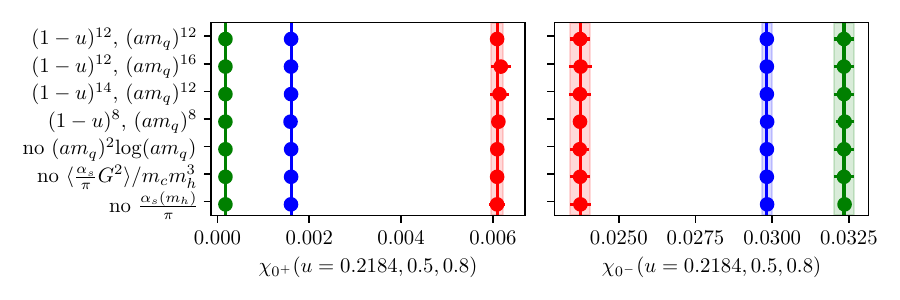}
\caption{\label{stability0m0p} Comparison of $\chi_{0^-}(u)$(left) and $\chi_{0^+}(u)$(right) at $u=0.2184$~(red), $u=0.5$~(blue) and $u=0.8$~(green) computed using the variations of the fit described in~\cref{stabsec}, indicated on the vertical axis. The topmost value and filled band correspond to our final results. We see that our results vary only very slightly for these different methods of performing the extrapolation.}
\end{figure*}

\begin{figure*}
\centering
\includegraphics[scale=1]{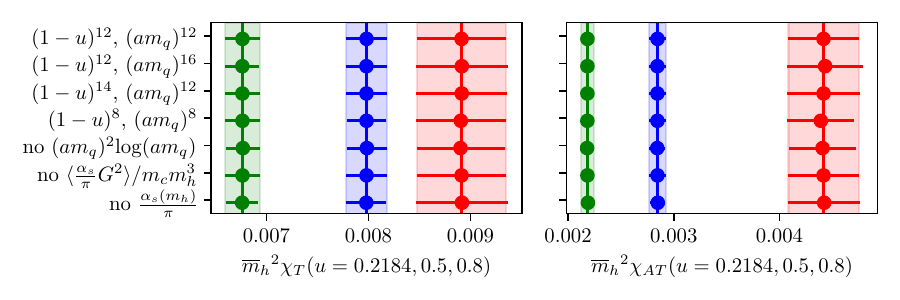}
\caption{\label{stabilityTAT} Comparison of $\chi_{T}(u)$(left) and $\chi_{AT}(u)$(right) at $u=0.2184$~(red), $u=0.5$~(blue) and $u=0.8$~(green) computed using the variations of the fit described in~\cref{stabsec}, indicated on the vertical axis. The topmost value and filled band correspond to our final results. We see that our results vary only very slightly for these different methods of performing the extrapolation.}
\end{figure*}

\cref{stability1m1p,stability0m0p,stabilityTAT} show the values of the susceptibilities at $u=0.2184$, $u=0.5$ and $u=0.8$ computed using the variations of the fit described in~\cref{stabsec}. We see that our results are insensitive to such variations in fitting strategy.

\end{appendix}

\bibliographystyle{apsrev4-1}
\bibliography{BsDsstar}

\end{document}